\documentclass{ws-procs975x65}

\begin{document}

\title{Unidentified EGRET sources
}

\author{G.~E. Romero
\footnote{\uppercase{M}ember of CONICET.}
}

\address{Instituto Argentino de Radioastronom\'{\i}a (IAR) \\
C.C. No. 5, 
1894, Villa Elisa, Buenos Aires, 
Argentina.\\ 
E-mail: romero@irma.iar.unlp.edu.ar\\ 
}


\maketitle

\abstracts{Some general properties of the unidentified EGRET sources are briefly discussed.}

The Compton Gamma-Ray Observatory was launched in early 1991 and deorbited in June, 2000. Between April 1991 and September 1995, the Energetic Gamma Ray Experiment Telescope (EGRET) detected 416 gamma-ray excesses with a significance of more than 3 $\sigma$ above the diffuse background emission. After a careful analysis with a maximum-likelihood procedure, only 271 of these sources were included in the final official list published as the Third EGRET (3EG) Catalog (Hartman et al. 1999). The included detections have a significance of more than 5 $\sigma$ at less than 10 degrees from the galactic plane, where the background radiation is stronger, and more than 4 $\sigma$ otherwise. Single power-law spectra have been fitted to most sources, and the corresponding spectral photon index is given in the catalog. These indices are usually between 1.5 and 3.0. The catalog also provides information on the location of the sources, in the form of the radius, in degrees, of a circle containing the same solid angle as the 95 \% confidence location contour of the gamma-ray excess. Integrated flux $F(E>100$ MeV) is provided for each viewing period in which the source was detected. The catalog also provides information on possible extended or complex morphology and potential identifications, when available. 

At the time the catalog was published, only a fraction of the sources was identified. Among these identifications we can mention 5 pulsars (i.e. rotating magnetized neutron stars where the pulsation was detected in the gamma-ray signal), 1 solar flare (a transient source), 1 normal galaxy other than ours (the Large Magallanic Cloud), one radio galaxy (Centaurus A), and several dozens of Active Galactic Nuclei (AGNs). Since then, only a few more AGNs and an additional pulsar have been added to the list. Most of the detected gamma-ray sources in the range 100 MeV -- 20 GeV are then unidentified objects.  


Significant efforts have been devoted in recent years to identify the large number of EGRET detections. These attempts are based on positional correlation studies with sources that are visible at lower energies in the EGRET location error boxes, on statistical population studies, on gamma-ray variability and spectral studies, and on multiwavelengths studies of special sources.


A simple analysis of the distribution of the unidentified EGRET sources with galactic coordinates reveals that there is a clear concentration of sources on the galactic plane plus a concentration in the general direction of the galactic center. This indicates a significant contribution from galactic sources. If many or most of the low-latitude gamma-ray sources are produced by galactic objects, one of the first things we could ask is whether they are correlated with the spiral arms of the Galaxy, i.e. the places where stars are formed. A correlation analysis between 3EG sources and bright and giant HII regions (the usual tracer for the galactic spiral structure) shows that there is a strong correlation at $\sim 7\sigma$-level (Romero 2001). This means that there is a significant number of Population I objects in the parent population of the low-latitude gamma-ray sources. 



If most of these low-latitude sources are young stellar-like objects, then it would be interesting to study the correlation between EGRET sources on the one hand, and young objects like massive stars, supernova remnants (SNRs), young pulsars, and OB associations, on the other. A number of such studies have been carried out in recent years (e.g. 
Yadigaroglu \& Romani 1997, Romero et al. 1999, Torres et al. 2002). They indicate that there is overwhelming statistical evidence supporting an association between EGRET sources and SNRs/star forming regions. The probability of these associations being a mere effect of chance is extremely low, below $10^{-5}$. This does not mean that necessarily most of the EGRET sources coincident with SNRs should be produced by the remnants themselves. It could be the case that a compact object also formed in the same SN explosion is responsible for the emission, or that since SNRs are usually found in star-forming regions, other young objects in the associations are generating the observed gamma-rays. 



Because of the huge error boxes in the location of the EGRET sources (typically $\sim 1$ degree in diameter), correlation studies have a limited potential to find secure counterparts. This technique has been successful mainly to identify bright blazars at high latitudes, where the confusion is significantly lower than in the galactic plane. A complementary approach are population studies using the known characteristics of the gamma-ray sources, like their spectra, variability and flux density. The so-called $\log S - \log N$ studies can be particularly useful in this respect. A $\log S - \log N$ plot displays the number of sources ($N$) with a gamma-ray flux ($S$) greater than a given value. By means of such an analysis of the steady sources in the 3EG catalog, Gehrels et al. (2000) have demonstrated the existence of at least two well-defined populations: one, a group of bright sources at low latitudes with an average spectral index of $2.18\pm0.04$, and the other a group of weaker mid-latitude sources with softer spectra (an average index of $2.49\pm0.04$). The transition between both groups occurs at $|b|\sim 5$ degrees, i.e. far from where the source significance changes from $4\sigma$ to $5\sigma$. The low-latitude sources are probably young objects located in the inner spiral arms, whereas the mid-latitude sources might be nearby (100-400 pc) sources associated to the Gould Belt, a local star forming region (Grenier 2001).

More recent and complete population studies including variability and source distribution models that take into account the non-uniform detection sensitivity across the sky (Grenier 2001, 2004) suggest that there exist at least 4 different populations of gamma-ray sources: 1) Bright and relatively hard (photon index $\Gamma\sim 2.5$) sources near the plane ($|b|<5^{\circ}$), some of them variable. 2) Weaker sources and non-variable sources with $\Gamma\sim2.25$ that are spatially correlated with the Gould Belt. 3) A group of luminous sources with very soft spectra ($\Gamma\sim2.5$) and high variability forming a halo around the galactic center with a scale height of $\sim2$ kpc. 4) An isotropic population of extragalactic origin and a variety of spectra and variability behaviors. There are no more than 35 sources in this group.

Regarding the galactic sources, the first group should be formed by young sources (a few million years at most) with isotropic luminosities in the range $10^{34-36}$ erg/s. The group 2 should be composed by also young sources but with luminosities in the range $10^{32-33}$ erg/s. The sources in the third group might be formed in and ejected from globular clusters (there is no significant correlation with individual clusters) or from the galactic plane. These sources should be old (age measured in Gyrs) and very luminous, in the range $10^{35-37}$ erg/s.    

The sources forming a given population do not need to be of homogeneous nature. For instance, the extragalactic sources can be different types of AGNs, the sources in the plane a mixture of young sources like SNRs, stellar systems, pulsars, microquasars, etc. In addition to the main populations there might be some number of rarer objects not so easy to classify, as discussed, by instance, by Punsly et al. (2000). 





\end{document}